\documentclass[submission,copyright,creativecommons]{eptcs}
\providecommand{\event}{TARK 2017} 
\usepackage{breakurl}             
\usepackage{underscore}           

\usepackage{amssymb}
\usepackage{graphicx}
\usepackage{framed}
\usepackage{tikz,subfigure}
\usepackage{amstext}
\usepackage{amssymb}
\usepackage{amsmath}
\usepackage{graphicx}
\usepackage{graphics} 
\usepackage[all,arc,poly]{xy}
\usepackage{color}
\usepackage{relsize}

\newcommand{\iel}{\textsf{IEL}}
\renewcommand{\phi}{\varphi}

\newcommand\<{\langle}
\renewcommand\>{\rangle}

\newcommand{\nc}{\newcommand}
\nc{\look}{\marginpar{$\bullet$}}
\nc{\Section}{\section}
\nc{\SubSection}{\subsection}

\newtheorem{theo}{Theorem}
\newtheorem{ddef}[theo]{Definition}
\newtheorem{proof}{Proof}
\newtheorem{llem}[theo]{Lemma} 
\newtheorem{oobs}[theo]{Observation} 
\newtheorem{rrem}[theo]{Remark} 
\newtheorem{prop}[theo]{Proposition} 
\newtheorem{ccor}[theo]{Corollary}  
\newtheorem{qquest}[theo]{Question} 
\newtheorem{fact}[theo]{Fact} 
\newtheorem{pprov}[theo]{Proviso}
\newtheorem{eexam}[theo]{Example} 

\nc{\bT}{\begin{theo}} 
\nc{\eT}{\end{theo}}
\nc{\bD}{\begin{ddef} \rm }
\nc{\eD}{\end{ddef}}
\nc{\bC}{\begin{ccor}}
\nc{\eC}{\end{ccor}}
\nc{\bCl}{\begin{claim}}
\nc{\eCl}{\end{claim}}
\nc{\bQ}{\begin{qquest}}
\nc{\eQ}{\end{qquest}}
\nc{\bL}{\begin{llem}}
\nc{\eL}{\end{llem}}
\nc{\bP}{\begin{prop}}
\nc{\eP}{\end{prop}}
\nc{\bR}{\begin{rrem}}
\nc{\eR}{\end{rrem}}
\nc{\bO}{\begin{oobs}}
\nc{\eO}{\end{oobs}}
\nc{\bF}{\begin{fact}}
\nc{\eF}{\end{fact}}
\nc{\bProv}{\begin{pprov}}
\nc{\eProv}{\end{pprov}}
\nc{\bE}{\begin{eexam} \rm }
\nc{\eE}{\end{eexam}}

\renewcommand{\geq}{\geqslant}

\renewcommand{\subset}{\subseteq}

\newcommand{\kbox}{\ensuremath{\square}}
\newcommand{\ibox}{\boxplus}
\def\disji{\rotatebox[origin=c]{-90}{$\!{\geqslant}$}}
\newcommand{\lori}{\,\disji\,}
\newcommand{\simn}{{\sim^n}}

\def\Disji{\rotatebox[origin=c]{-90}{$\!{\mathlarger{\mathlarger{\,\mathlarger{\geqslant}}}}\!$}}
\newcommand{\Lori}{\,\Disji\,}
\newcommand{\ee}{\epsilon}
\newcommand{\s}{\texttt{s}}
\newcommand{\us}{\underline{\texttt{s}}}
\newcommand{\E}{\texttt{E}}

\newcommand{\PP}{\texttt{P}}

\nc{\simg}{\sim_{\mathsf{g}}}
\nc{\ssimg}{\approx_{\mathsf{g}}}
\nc{\FO}{\mathsf{FO}}
\nc{\MSO}{\mathsf{MSO}}
\nc{\ML}{\mathsf{ML}}
\nc{\IML}{\mathsf{IML}}
\nc{\SO}{\mathsf{SO}}
\nc{\GF}{\mathsf{GF}}
\nc{\gd}{\mathsf{gd}}
\nc{\hg}{\mathsf{hg}}
\nc{\vg}{\mathsf{vg}}
\nc{\free}{\mathrm{free}}
\nc{\dom}{\mathrm{dom}}
\nc{\nn}{\mathsf{non}}
\nc{\NE}{\mathsf{NE}}
\nc{\NF}{\mathsf{NF}}

\nc{\full}{\mathsf{full}}
\nc{\lf}{\mathsf{lf}}
\nc{\dc}{\mathsf{dc}}
\nc{\rel}{\mathsf{rel}}

\newenvironment{romanenumerate}%
{\begin{list}{(\roman{enumi})}{\usecounter{enumi}
\setlength{\labelwidth}{2cm}
\setlength{\itemindent}{0pt}
\setlength{\itemsep}{0.5\itemsep}
\setlength{\topsep}{\itemsep}
\setlength{\parsep}{0pt}
}}{\end{list}}
\nc{\bre}{\begin{romanenumerate}}
\nc{\ere}{\end{romanenumerate}}
\newenvironment{alphaenumerate}%
{\begin{list}{(\alph{enumii})}{\usecounter{enumii}
\setlength{\labelwidth}{2cm}
\setlength{\itemindent}{0pt}
\setlength{\itemsep}{0.5\itemsep}
\setlength{\topsep}{\itemsep}
\setlength{\parsep}{0pt}
}}{\end{list}}
\nc{\bae}{\begin{alphaenumerate}}
\nc{\eae}{\end{alphaenumerate}}
\newenvironment{numenumerate}%
{\begin{list}{(\arabic{enumiii})}{\usecounter{enumiii}
\setlength{\labelwidth}{2cm}
\setlength{\itemindent}{0pt}
\setlength{\itemsep}{0.5\itemsep}
\setlength{\topsep}{\itemsep}
\setlength{\parsep}{0pt}
}}{\end{list}}
\nc{\bne}{\begin{numenumerate}}
\nc{\ene}{\end{numenumerate}}

\nc{\ins}[1]{\bigskip\noindent
\framebox{\begin{minipage}{.47\textwidth} \sloppy \noindent \em #1 \end{minipage}}\bigskip}

\nc{\str}[1]{{\mathfrak{#1}}}
\nc{\brck}[1]{[\![ #1 ]\!]}
\nc{\restr}{\!\restriction\!}

\nc{\HH}{\mathbb{H}}
\nc{\VV}{\mathbb{V}}
\nc{\abar}{\mathbf{a}}
\nc{\bbar}{\mathbf{b}}
\nc{\cbar}{\mathbf{c}}
\nc{\xbar}{\mathbf{x}}
\nc{\ybar}{\mathbf{y}}
\nc{\zbar}{\mathbf{z}}
\nc{\ubar}{\mathbf{u}}
\nc{\sbar}{\mathbf{s}}
\nc{\tbar}{\mathbf{t}}
\nc{\vbar}{\mathbf{v}}
\nc{\wbar}{\mathbf{w}}

\nc{\Xbar}{\mathbf{X}}
\nc{\Ybar}{\mathbf{Y}}
\nc{\Zbar}{\mathbf{Z}}
\nc{\Pbar}{\mathbf{P}}
\nc{\nubar}{\mbox{\boldmath $\nu$}}

\nc{\barr}{\begin{array}}
\nc{\earr}{\end{array}}
\nc{\btab}{\begin{tabular}}
\nc{\etab}{\end{tabular}}

\nc{\nothing}{\rule{0em}{1ex}}
\nc{\highnothing}{\rule{0em}{3ex}}
\nc{\hnt}{\highnothing}
\nc{\nt}{\nothing}
\nc{\nnt}{\rule{.1pt}{0pt}}

\nc{\ssc}{\scriptscriptstyle}
\nc{\N}{{\mathbb N}}
\nc{\Z}{{\mathbb Z}}
\nc{\M}{{\mathbb M}}
\nc{\F}{{\mathbb F}}
\nc{\W}{{\mathbb W}}
\newcommand{\PI}{\mbox{\bf I}}
\newcommand{\PII}{\mbox{\bf II}}
\renewcommand{\P}{\ensuremath{\mathcal{P}}}

\newcommand{\CC}{\ensuremath{\mathcal{C}}}

\newcommand{\MM}{\ensuremath{\mathfrak{M}}}
\renewcommand{\epsilon}{\varepsilon}

\newcommand{\inqbm}{\textsc{InqML}}
\newcommand{\inqml}{\textsc{InqML}}

\nc{\prf}{\begin{proof}}
\nc{\eprf}{\end{proof}}

\title{Bisimulation in Inquisitive Modal Logic\thanks{Ivano Ciardelli's research was financially supported by the European Research Council (ERC) under the European Union's Horizon 2020 research and innovation programme (grant agreement No. 680220). Martin Otto's research was partially funded by DFG grant OT 147/6-1:
\emph{Constructions and Analysis in Hypergraphs of Controlled Acyclicity.}}}
\author{Ivano Ciardelli
\institute{Institute for Logic, Language, and Computation\\ University of Amsterdam}
\email{i.a.ciardelli@uva.nl}
\and
Martin Otto
\institute{Department of Mathematics, Logic Group \\ Technische Universit\"at Darmstadt}
\email{otto@mathematik.tu-darmstadt.de}
}
\def\titlerunning{Bisimulation in Inquisitive Modal Logic}
\def\authorrunning{I. Ciardelli and M. Otto}
\begin{document}
\maketitle

\begin{abstract}
Inquisitive modal logic, \inqml, is a generalisation of standard Kripke-style modal logic. In its epistemic incarnation, it extends standard epistemic logic to capture not just the information that agents have, but also the questions that they are interested in. Technically, \inqml\ fits within the family of logics based on team semantics. From a model-theoretic perspective, it takes us a step in the direction of monadic second-order logic, as inquisitive modal operators involve quantification over sets of worlds. We introduce and investigate the natural notion of bisimulation equivalence in the setting of \inqml. We compare the expressiveness of \inqml\ and first-order logic, and characterise inquisitive modal logic  as the bisimulation invariant fragments of first-order logic over various classes of two-sorted relational structures. These results crucially require non-classical methods in studying bisimulations and first-order expressiveness over non-elementary classes.\end{abstract}

\section{Introduction}

\noindent The recently developed framework of \emph{inquisitive logic} \cite{CiardelliRoelofsen,IvanoDiss,Ciardelli:15inqd,Ciardelli:16qait} can be seen as a generalisation of classical logic which encompasses not only statements, but also questions. One reason why this generalisation is interesting is that it provides a novel perspective on the logical notion of \emph{dependency}, which plays an important r\^{o}le in applications (e.g., in database theory) and which has recently received attention in the field of \emph{dependence logic} \cite{Vaananen:07}. 
Indeed, dependency is nothing but a facet of the fundamental logical relation of entailment, once this is extended so as to apply not only to statements, but also to questions \cite{Ciardelli:16dependency}. This connection explains the deep similarities existing between systems of inquisitive logic and systems of dependence logic (see \cite{Yang:14,Ciardelli:16dependency,IvanoDiss,YangVaananen}).
A different r\^{o}le for questions in a logical system comes from the setting of modal logic: once the notion of a modal operator is suitably generalised, questions can be embedded under modal operators to produce new statements that have no ``standard'' counterpart. This approach was first developed in \cite{CiardelliRoelofsen:15idel} in the setting of epistemic logic. The resulting \emph{inquisitive epistemic logic} (\iel) models not only the information that agents have, but also the issues that they are interested in, i.e., the information that they would like to obtain. Modal formulae in \iel\ can express not only that an agent knows that~$p$ ($\kbox p$) but also that she knows \emph{whether}~$p$ ($\kbox{?p}$) or that she \emph{wonders whether}~$p$ ($\ibox{?p}$)---a statement that cannot be expressed without the use of embedded questions. As shown in \cite{CiardelliRoelofsen:15idel}, several key notions of epistemic logic generalise smoothly to questions: besides common knowledge we now have \emph{common issues}, the issues publicly entertained by the group; and besides publicly announcing a statement, agents can now also publicly ask a question, which typically results in new common issues. 
Thus, \iel\ may be seen as one step in extending modal logic 
from a framework to reason about information and information change, 
to a richer framework which also represents a higher stratum of cognitive phenomena, in particular issues and their raising in a communication scenario. 

Of course, like standard modal logic, inquisitive modal logic provides a general framework that admits various
 interpretations, each suggesting corresponding constraints on models.
E.g., \cite{IvanoDiss} suggests to interpret \inqml\ as a logic of action. On this interpretation, a modal formula $\kbox{?p}$ 
expresses that whether a certain fact $p$ will come about is determined independently
of the agent's choices, while $\ibox{?p}$ expresses that
whether  $p$ will come about is fully determined by her choices.

From the perspective of mathematical logic, inquisitive modal logic is a natural
generalisation of standard modal logic. There, the accessibility
relation of a Kripke model associates 
with each possible world $w\in W$ a set $\sigma(w) \subset W$ of
possible worlds, namely, the worlds accessible from $w$; 
any formula $\phi$ of modal logic is semantically
associated with a set $|\phi|_\M \subset W$ of worlds, namely, the set of worlds where it is true; modalities then express relationships between these sets: 
for instance, $\kbox\phi$ expresses the fact that $\sigma(w)\subseteq|\phi|_\M$.  
In the inquisitive setting, the situation is analogous, but both the
entity $\Sigma(w)$ attached to a possible world and the semantic extension
$[\phi]_\M$ of a formula are sets \emph{of sets} of
worlds, rather than simple sets of worlds. Inquisitive modalities still express
relationships between 
these two objects: $\kbox\phi$ expresses the fact that
$\bigcup\Sigma(w)\in[\phi]_\M$, while $\ibox\phi$ 
expresses the fact that $\Sigma(w)\subseteq[\phi]_\M$.

In this manner, inquisitive logic leads to a new framework for modal
logic that can be viewed as a generalisation of the standard framework. 
Clearly, this raises the question of whether and how the classical
notions and results of modal logic carry over to this more general
setting. In this paper we address this question for the fundamental notion 
of \emph{bisimulation} and for two classical results revolving around
this notion, namely, the Ehrenfeucht-Fra\"iss\'e theorem for modal 
logic, and van Benthem style characterisation
theorems~\cite{GorankoOtto,Benthem83,Rosen,OttoNote}. 
A central topic of this paper is the r\^ole of \emph{bisimulation invariance} 
as a unifying semantic feature that distinguishes modal logics
from classical predicate logics. As in many other areas, from 
temporal logics and process logics to knowledge representation in AI 
and database applications, so also in the inquisitive setting 
we find that the appropriate notion of  bisimulation invariance allows for 
precise model-theoretic characterisations of the expressive power of modal logic 
in relation to first-order logic.

Our first result is that the right notion of
inquisitive bisimulation equivalence $\sim$, with finitary approximation levels 
$\simn$, 
supports a counterpart for \inqbm\ of 
the classical  Ehrenfeucht--Fra\"\i ss\'e correspondence.
This result is non-trivial 
in the inquisitive setting, because of some subtle issues
stemming from the interleaving of first- and second-order features
in inquisitive modal logic. 

\medskip
\bT[inquisitive Ehrenfeucht--Fra\"\i ss\'e theorem]~\\
\label{EFthm} 
Over finite vocabularies, the finite levels $\simn$ of 
inquisitive bisimulation equivalence correspond to 
the levels of $\inqbm$-equivalence up to modal nesting depth $n$. 
\label{mainEF}
\eT



In order to compare \inqbm\ with classical first-order logic, we define a class
of two-sorted relational structures, and show how such structures encode 
models for \inqbm.  With respect to such relational structures
we find not only a ``standard translation'' of \inqbm\   
into two-sorted first-order logic, but also a van Benthem style
characterisation of $\inqbm$ as the bisimulation-invariant fragment of 
(two-sorted) first-order logic over several classes of models. These results are technically
interesting, and they are not available on the basis of classical techniques,
because the relevant classes of two-sorted models are 
non-elementary (in fact, first-order logic is not compact over these classes, as we show). 
Our techniques yield characterisation theorems both in the setting of arbitrary 
 inquisitive models, and in restriction to just finite ones. 

\bT
\label{main1}
Inquisitive modal logic can be characterised as the 
$\sim$-invariant fragment of first-order logic $\FO$ over natural 
classes of (finite or arbitrary) relational inquisitive models. 
\eT

%


\noindent
Beside the conceptual development and the core results themselves, we think that 
also the methodological aspects of the present investigations have
some intrinsic value. Just as inquisitive logic models cognitive phenomena  
at a level strictly above that of standard modal logic, so the
model-theoretic analysis moves up from the level of ordinary first-order logic 
to a level strictly between first- and second-order logic. This level
is realised by first-order logic in a two-sorted framework that
incorporates second-order objects in the second sort in a controlled fashion. 
This leads us to substantially generalise
a number of notions and techniques developed in the model-theoretic analysis of modal logic
(\cite{GorankoOtto,OttoNote,DawarOttoAPAL09,OttoAPAL04}, among others).
\section{Inquisitive modal logic}
\label{sec:inquisitive modal logic}

\noindent 
In this section we provide an essential introduction to inquisitive
modal logic, \inqbm\ \cite{IvanoDiss}.
For details and proofs, see \S7 of \cite{IvanoDiss}.

\subsection{Foundations of Inquisitive Semantics}

\noindent
Usually, the semantics 
of a logic specifies truth-conditions for the formulae of the logic. 
In modal logics these truth-conditions are relative to possible worlds
in a Kripke model. 
However, this approach is limited in an important way: while suitable for statements, 
it is inadequate for questions. To overcome this limitation, inquisitive logic
interprets formulae not relative to states of affairs (worlds), but
relative to states of information, 
modelled extensionally as sets of worlds (viz., those worlds compatible with the given~information).

\smallskip
\begin{ddef}[information states] 
An \emph{information state} over a set of worlds $W$ is a subset
$s\subseteq W$. 
\end{ddef}

Rather than specifying when a sentence is \emph{true} at a world $w$, 
inquisitive semantics specifies when a sentence is \emph{supported} by an information state $s$: for a statement $\alpha$ this means that the information available in
$s$ implies that $\alpha$ is true; for a question~$\mu$, it means that
the information available in $s$ settles~$\mu$.
%
%
If $t$ and $s$ are information states and $t\subseteq s$, this means that
$t$ holds at least as much information as $s$: we say that $t$ is an \emph{extension} of~$s$. If $t$ is an
extension of $s$, everything that is supported at $s$ will also
be supported at $t$. This is a key feature of inquisitive
semantics, and it leads naturally to the notion of an \emph{inquisitive state}
(see \cite{Ciardelli:13compass,Roelofsen:13,CiardelliRoelofsen:15idel}).

\smallskip
\begin{ddef}[inquisitive states]~\\
An \emph{inquisitive state} over 
$W$ is a 
non-empty set of information states $\Pi\subseteq\wp(W)$ satisfying
\begin{itemize}
\item
$s\in\Pi$ and $t\subseteq s$ implies  $t \in\Pi$ (downward closure).
\end{itemize}
\end{ddef}







\subsection{Inquisitive Modal Models}

\noindent A Kripke frame can be thought of as a set $W$ of worlds 
together with a map $\sigma$ that equips each world with a set of 
worlds $\sigma(w)$---the set of worlds that are accessible from
$w$---i.e., an information state.

Similarly, an inquisitive modal frame consists of a set $W$ of worlds
together with an \emph{inquisitive assignment}, i.e., a map $\Sigma$
that assigns to each world an inquisitive state.
An inquisitive modal model is an inquisitive frame equipped with a
valuation function.

\medskip
\begin{ddef}[inquisitive modal models]~\\
An inquisitive modal frame is a pair $\F=\<W,\Sigma\>$, where
$\Sigma\colon W\to\wp\wp(W)$ associates  to each world $w\in W$ an inquisitive state $\Sigma(w)$.
An inquisitive modal model is a pair $\M=\<\F,V\>$ where $\F$ is an
inquisitive modal frame, and $V \colon \P \rightarrow \wp(W)$
is a propositional valuation function.
A {world-(or state-)pointed} inquisitive modal model is a pair
consisting of a model $\M$ and a distinguished world (or state) in $\M$.
\end{ddef}

\noindent 
With an inquisitive modal model $\M$ we can always associate a
standard Kripke model $\str{K}(\M)$ having the same set of worlds and  
modal accessibility map $\sigma:W\to\wp(W)$ induced by the inquisitive
map $\Sigma$ according to $\sigma(w):=\bigcup\Sigma(w)$. 
%

Under an epistemic interpretation \cite{CiardelliRoelofsen:15idel,Ciardelli:14aiml}, $\Sigma$ is taken to describe not only an agent's \emph{knowledge}, as in epistemic logic, but also her \emph{issues}, i.e., the questions she is interested in. The agent's knowledge state at $w$, $\sigma(w)=\bigcup\Sigma(w)$, consists of those worlds that are compatible with what the agent knows. The agent's inquisitive state at $w$, $\Sigma(w)$, consists of those information states where her issues are settled. 

%
%

\subsection{Inquisitive Modal Logic}

\noindent
The syntax  of inquisitive modal logic $\inqbm$ is given by:
$$\phi::= p\;|\,\bot\,|\,(\phi\land\phi)\,|\,(\phi\to\phi)\,|\,(\phi\lori\phi)\,|\,\kbox\phi\,|\,\ibox\!\phi$$
%

We treat negation and disjunction as defined connectives (syntactic
shorthands) according to
$\neg\phi:=\phi\to\bot$, and $\phi\lor\psi:=\neg(\neg\phi\land\neg\psi)$. 
In this sense, the above syntax includes standard propositional formulae
in terms of atoms and connectives $\land$ and $\to$ together with the defined
$\neg$ and $\vee$. As we will see, the semantics for such formulae will be
essentially the same as in standard propositional logic. 
%
%
%
In addition to standard connectives, our language contains a new
connective, $\lori$, called \emph{inquisitive disjunction}. We may read
formulae built up by means of this connective as propositional
questions. E.g., we read the formula $p\lori\neg p$ as the question
\emph{whether or not $p$}, and we abbreviate this formula as $?p$.  Finally, our language contains two modalities, which are allowed to
embed both statements and questions. 
As we shall see, both these modalities coincide with a standard Kripke
box when applied to statements, but crucially differ when applied to questions. 
Under an epistemic interpretation, $\kbox{?p}$ expresses 
the fact that the agent knows whether $p$, while $\ibox{?p}$ expresses
(roughly) the fact that she wants to find out whether~$p$. 

While models for \inqbm\ are formally a class of neighbourhood models, 
the semantics of \inqbm\ is very different from neighbourhood
semantics for modal logic.\footnote{This different perspective on the
  models leads us to a notion of bisimulation which is different
  from the one that has been considered for neighbourhood models \cite{Hansen:03,Pacuit:07,Hansen:09}. Due to space limitations we cannot discuss the difference in detail here.} 
As mentioned above, the semantics of \inqbm\ is given in terms of support relative to an information state, rather than truth at a possible world. 


\begin{ddef}[semantics of $\inqbm$]~\\
\label{supportsemdefn}%
Let $\M=\<W,\Sigma,V\>$ be an inquisitive modal model, $s\subseteq W$:
\begin{itemize}
\item $\M,s\models p\iff s\subseteq V(p)$
\item $\M,s\models \bot\iff s=\emptyset$
\item $\M,s\models \phi\land\psi\iff \M,s\models\phi\text{ and }\M,s\models\psi$
\item $\M,s\models \phi\to\psi\iff \forall t\subseteq s:\M,t\models\phi\Rightarrow \M,t\models\psi$
\item $\M,s\models \phi\lori\psi\iff \M,s\models\phi\text{ or }\M,s\models\psi$
\item $\M,s\models \kbox\phi\iff\forall w\in s: \M,\sigma(w)\models\phi$
\item $\M,s\models \ibox\phi\iff\forall w\in s\;\forall t\in\Sigma(w): \M,t\models\phi$
\end{itemize}
\end{ddef}

\noindent
As an illustration, consider the support conditions for the formula $?p:=p\lori\neg p$: this formula is supported by a state
$s$ in case $p$ is true at all worlds in $s$ (i.e., if the information available in $s$ implies that $p$ is true)
or in case $p$ is false at all worlds in $s$ (i.e., if the information available in $s$ implies that $p$ is false).
Thus, $?p$ is supported precisely by those information states that settle whether 
or not $p$ is true.

The following two properties hold generally in \inqbm: 
\begin{itemize}
\item Persistency: if $\M,s\models\phi$ and $t\subseteq s$, then $\M,t\models\phi$;
\item Semantic ex-falso: $\M,\emptyset\models\phi$ for all $\phi\in\inqbm$.
\end{itemize}

\noindent
The first principle says that support is
preserved as information increases, i.e., as we move from a state to an extension of it. The second principle says
that the empty set of worlds---the inconsistent state---vacuously supports everything. 
Together, these principles imply that the support set $[\phi]_\M :=
\{ s \subseteq W \colon \M,s \models \phi \}$ of a formula is downward closed and non-empty, i.e.,  
it is an inquisitive state.


Although the primary notion of our semantics is support at an information state, truth at a world is obtained as a defined notion.

\begin{ddef}[truth] 
\label{truthdef}
 $\phi$ is \emph{true} at a world $w$ in a model $\M$, 
denoted $\M,w\models\phi$, in case $\M,\{w\}\models\phi$.
\end{ddef}

%
%
%
%
%

Spelling out Definition~\ref{truthdef} in the special case of
singleton states, we see that standard connectives have the usual
truth-conditional behaviour. 
For modal formulae, we find the following truth-conditions:

\medskip
\begin{prop}[truth conditions for modal formulae]~
\begin{itemize}
\item $\M,w\models \kbox\phi\iff \M,\sigma(w)\models\phi$
\item $\M,w\models \ibox\phi\iff \forall t\in\Sigma(w): \M,t\models\phi$
\end{itemize}
\end{prop}

\noindent 
Notice that truth in \inqbm\ cannot be given a direct recursive definition, as the truth conditions for modal formulae $\kbox\phi$ and $\ibox\phi$ depend on the support conditions for $\phi$---not just on its truth conditions.

 For many formulae, support at a state just boils down to truth at each world. We refer to these formulae as \emph{truth-conditional}.
 \footnote{In dependence logic (e.g., \cite{Vaananen:07,YangVaananen}) truth-conditional formulae are called \emph{flat} formulae.}

\medskip
 \begin{ddef}[truth-conditional formulae]\label{flatdef}
 We say that a formula $\phi$ is \emph{truth-conditional} if for all models $\M$ and information states $s$: $\M,s\models\phi\iff \M,w\models\phi$ for all $w\in s$.
 \end{ddef}

Following \cite{IvanoDiss}, we view truth-conditional 
formulae as statements, and non-truth-conditional formulae as
questions. The next proposition identifies a large class of truth-conditional formulae.


\begin{prop}\label{prop:truth-conditionality} \mbox{}
\label{flatprop}
Atomic formulae, $\bot$, 
and all formulae of the form $\kbox\phi$ and $\ibox\phi$ are truth-conditional.
The class of truth-conditional formulae is closed under all
connectives except for $\lori$.
\end{prop}

Using this fact, it is easy to see that all formulae of standard modal logic, i.e., formulae which do not contain $\lori$ or $\ibox$, receive exactly the same truth conditions as in standard modal logic.

\begin{prop} 
If $\phi$ is a formula not containing $\lori$ or $\ibox$, then $\M,w\models\phi\iff\str{K}(\M),w\models\phi$ in standard Kripke semantics.
\end{prop}

As long as questions are not around, the modality~$\ibox$ also coincides with $\kbox$, and with the standard box modality. That is, if $\phi$ is truth-conditional, we have:
$$M,w\models\kbox\phi\iff M,w\models\ibox\phi\iff M,v\models\phi\text{ for all }v\in\sigma(w)$$


Thus, the two modalities coincide on statements. However, they come apart when they are applied to questions. For an illustration, consider the formulae $\kbox{?p}$ and $\ibox{?p}$ in the epistemic setting: $\kbox{?p}$ is true iff the information state of the agent, $\sigma(w)$, settles the question $?p$; thus, $\kbox{?p}$ expresses the fact that the agent knows whether $p$. By contrast, $\ibox{?p}$ is true iff any information state $t\in\Sigma(w)$, i.e., any state that settles the agent's issues, also settles $?p$; thus $\ibox{?p}$ expresses that finding out whether $p$ is part of the agent's goals.


%

\section{Inquisitive Bisimulation}
\label{sec:bisimulations}



\noindent
An inquisitive modal model can be seen as a structure with two sorts
of entities, worlds and information states, which interact with each
other. On the one hand, an information state $s$ is completely
determined by the worlds that it contains; on the other hand, a world
$w$ is determined by the atoms it makes true and the information
states which lie in $\Sigma(w)$. Taking a more behavioural
perspective, we can look at an inquisitive modal model as a model
where two kinds of transitions are possible: from an information state
$s$, we can make a transition to a world $w\in s$, and from a world
$w$, we can make a transition to an information state $s\in\Sigma(w)$. 
This suggests a natural notion of bisimilarity, together with its natural
finite approximations of $n$-bisimilarity for $n \in \N$. As usual, these
notions can equivalently be defined either in terms of back-and-forth
systems or in terms of strategies in corresponding bisimulation games. 
We chose the latter for its more immediate and intuitive appeal to the 
underlying dynamics of a ``probing'' of behavioural equivalence. 


The game is played by two players, \PI\ and \PII, 
%
%
who act as challenger and defender of a similarity
claim involving a pair of worlds $w$ and $w'$ or 
information states $s$ and $s'$ over two models
$\M=\langle W,\Sigma,V \rangle$ and $\M'=\langle W',\Sigma',V'\rangle$.
We denote world-positions as $\langle w,w' \rangle$ and 
state-positions as $\langle s,s' \rangle$, where $w \in W, w' \in W'$
and $s \in \wp(W), s' \in \wp(W')$, respectively.
The game proceeds in rounds that alternate between world-positions and
state-positions. 
Playing from a world-position $\langle w,w' \rangle$,
  \PI\ chooses an information state in the inquisitive state associated to
 one of these worlds ($s \in \Sigma(w)$ or 
$s' \in \Sigma'(w')$)  
and \PII\ must respond by choosing an information state on the
opposite side,
which results in a state-position $\langle s,s' \rangle$.  
Playing from a state-position
$\langle s,s'\rangle$,
  \PI\ chooses a world in either state ($w \in s$ or $w' \in s'$)
and \PII\ must respond by choosing a world from the other state,
which results in a world-position $\langle w,w' \rangle$.
A round of the game consists 
of four moves leading from a world-position to another. 


In the bounded version of the game, the number of rounds is fixed
in advance. In the unbounded version, the game is allowed to go on
indefinitely. Either player loses when stuck for a move. The game ends
with a loss for \PII\ in any world-position $\langle w,w'\rangle$ that shows a
discrepancy at the atomic level, i.e., such that $w$ and $w'$ disagree on the truth of some $p\in\P$. 
All other plays, including infinite runs of the unbounded game, are won by $\PII$.


\begin{ddef}[bisimulation equivalence]
Two world-pointed models $\M,w$ and $\M',w'$ are \emph{$n$-bisimilar}, 
$\M,w\,\simn\, \M',w'$, if \PII\ has a winning strategy in the $n$-round 
game starting from $\langle w,w'\rangle$.
$\M,w$ and $\M',w'$ are \emph{bisimilar}, denoted 
$\M,w\sim \M',w'$, if \PII\ has a winning strategy in the unbounded
 game starting from $\langle w,w'\rangle$. 
 
%
Two state-pointed models $\M,s$ and $\M',s'$ are ($n$-)bisimilar, denoted $\M,s\!\sim\!\M',s'$
(or $\M,s\,\simn\,\M',s'$), if every world in $s$
is ($n$-)bisimilar to some world in $s'$ and vice
versa.  

%
%
%
Two 
models $\M$ 
and $\M'$ are \emph{globally bisimilar}, denoted 
$\M\sim \M'$, if
every world in $\M$ is bisimilar 
to some world in $\M'$ 
and vice versa.
%
\end{ddef}



\section{An Ehrenfeucht--Fra\"\i ss\'e theorem} 

\noindent
The
crucial r\^ole of these notions of equivalence for the model
theory of inquisitive modal logic is brought out 
in a corresponding Ehrenfeucht--Fra\"\i ss\'e theorem.

Using the standard notion of the modal depth of a formula, 
we denote as $\inqbm_n$ the class of $\inqbm$-formulae of depth up to
$n$. 
It is easy to see that the
semantics of any formula in $\inqbm_n$
is preserved under $n$-bisimilarity; as a consequence, all of inquisitive modal logic is
preserved under full bisimilarity. The following analogue
of the classical Ehrenfeucht--Fra\"\i ss\'e theorem shows that, for finite sets \P\ of basic propositions, 
{$n$-bisimilarity} coincides with logical 
indistinguishability in $\inqbm_n$, which we denote as 
$\equiv^n_\inqbm$: 
\[
\M,s \equiv^n_\inqbm \M',s' \overset{\text{def}}{\iff} \left\{\barr{@{\;}r@{}}
\M,s \models \phi\;\;\Leftrightarrow\;\;
\M',s' \models \phi
\\
\mbox{for all } \phi \in \inqbm_n.\!\!\nt \earr\right.
\]

\medskip\noindent
\textbf{Theorem 1.} 
\emph{Given a finite set of atomic propositions \P, 
for any $n \in \N$ and inquisitive state-pointed modal models $\M,s$ and $\M',s'$:}
\[ \M,s\;\simn\; \M',s' \iff
\M,s \equiv^n_\inqbm\M',s'
\]


Notice that, by taking $s$ and $s'$ to be singleton states, we obtain
the corresponding connection for world-pointed models as a special case:
$\M,w\;\simn\; \M',w'\!\iff\! \M,w\equiv^n_\inqbm\M',w'$.
%
As customary, the crucial
implication of the theorem, from right to left, follows from the existence of
\emph{characteristic formulae} that define $\sim^n$-classes of
worlds, information states and inquisitive states over 
models---and it is here that the
finiteness of \P\ is crucial.

\bP[characteristic formulae for $\simn$-classes]
\label{prop:characteristic formulae}~\\  
For any world-pointed model $\M,w$ over a 
finite set of atomic propositions \P, 
and for any $n \in \N$
there is a formula $\chi^n_{\M,w} \in\inqbm$ of modal depth $n$ 
s.t.\ $\M',w' \models \chi^n_{\M,w} \iff
\M',w'\, \simn\, \M,w$. 
\eP

\medskip\noindent
\textbf{Proof. }
By simultaneous induction on $n$, we define formulae 
$\chi^n_{\M,w}$ together with auxiliary
formulae $\chi^n_{\M,s}$ and $\chi^n_{\M,\Pi}$ 
for all worlds $w$, information states $s$ and inquisitive
states $\Pi$ over $\M$. Given two inquisitive states $\Pi$ and $\Pi'$
in models $\M$ 
and $\M'$, 
we write $\M,\Pi \sim^n \M',\Pi'$ if every  
state $s\in \Pi$ is $n$-bisimilar to some 
state $s'\in\Pi'$, and vice versa. 
%
Dropping reference to the fixed $\M$, 
we let:
\begin{eqnarray*}\label{def:characteristic formulae}
\chi^0_w\!\!\!\!&=&\!\!\!\!\bigwedge
\{p\colon w\in V(p) \}
\land\bigwedge
\{\neg p\colon w\not\in V(p) \}
\\
\chi^n_s\!\!\!\!&=&\!\!\!\!\bigvee\{\chi^n_w\colon w\in s\}
\\[.1cm]
\chi^n_\Pi\!\!\!\!&=&\!\!\!\!\Lori\{\chi^n_s \colon s\in\Pi\}
\\
\hnt\chi^{n+1}_w\!\!\!\!&=&\!\!\!\!\chi^n_w\land\ibox\chi^n_{\Sigma(w)}\!\land\bigwedge\{\neg\!\ibox\!\chi^n_\Pi\colon
                            \Pi\subseteq\Sigma(w),\, 
\Pi\not\!\!{\simn}\Sigma(w)\}
\end{eqnarray*}

These formulae are of the required modal depth; the conjunctions and
 disjunctions in the definition are well-defined since, for a
 given $n$, there are only finitely many distinct formulae of the
 form $\chi^n_w$, and analogously for $\chi^n_s$ or $\chi^n_\Pi$. We
 can then prove by simultaneous induction on $n$ that these formulae
 satisfy the following properties:
\begin{enumerate}
\item $\M',w'\models\chi^n_{\M,w}\iff \M',w'\,\simn\, \M,w$
\item $\M',s'\models\chi^n_{\M,s} \iff \M',s'\,\simn\, \M,t$ for some  $t\subseteq s$ 
\item $\M',s'\models\chi^n_{\M,\Pi}\iff \M',s'\,\simn\, \M,s$ for some $s\in\Pi$
\end{enumerate}
The details of the inductive proof are given in appendix \ref{EFappendixsec}.\hfill$\Box$\medskip
Let us say that a class \CC\ of world-pointed (state-pointed) models is \emph{defined}
by a formula $\phi$ if \CC\ is the set of world-pointed models where
$\phi$ is true (in which $\phi$ is supported).

\bC 
\label{EFcorrworldpointed}\label{EFcorrstatepointed} 
A class \CC\ of world-pointed models is definable in \inqbm\ if and only if it is closed under $\simn$ for some $n\in\N$.
A class \CC\ of state-pointed models is definable in \inqbm\ if and only if it is both downward closed and closed under $\simn$ for some $n\in\N$.
\eC







\section{Relational inquisitive models}
\label{relinqmodsec}


\noindent In this paper, we want to compare the expressive power of
inquisitive modal logic with that of first-order logic.  However, this
is not straightforward. 
A standard Kripke model can be identified naturally with a relational structure
with a binary accessibility relation $R$ 
and a 
unary predicate $P_i$  for the interpretation of each atomic sentence $p_i \in \P$. 
By contrast, an inquisitive modal model also needs to encode the
inquisitive state map $\Sigma:W\to\wp\wp(W)$. This map can be identified with a binary relation 
$E \subseteq W\times \wp(W)$. In order to view this as part of a relational structure, however, we need to adopt a two-sorted perspective, and view $W$ and $\wp(W)$ as domains of two distinct sorts.
%
%
%
This leads  to the following notion.

\subsection{Relational Inquisitive Models}

\begin{ddef}[relational models]\label{def:relational models}
A \emph{relational inquisitive modal model} is a relational structure 
\[
\str{M}=\<W,S,E,\ee,(P_i)_{p_i \in \P}\>
\]
where $W,S\not=\emptyset$ are sets, $E,\ee\subseteq W\times S$, and
$P_i\subseteq W$. With $\s\in S$ we associate the set 
$\us:=\{w\in W\,|\,w\ee\s\} \subset W$ and require the following conditions, which enforce resemblance with inquisitive modal models:
\begin{itemize}
\item Extensionality: if $\us=\us'$, then $\s=\s'$.
\item Non-emptiness: for every $w$, $E[w]\neq\emptyset$.
\item Downward closure: if $\s\in E[w]$ and $t\subseteq\us$, there is an $\s'\in S$ such that $\us'=t$ and $\s'\in E[w]$.
\end{itemize}
\end{ddef}

By extensionality, the second sort $S$ can be identified 
with a domain $\{\us\,|\,\s\in S\}
\subset \wp(W)$ of sets over the first sort. 
We will always make this identification and view a 
relational model as a structure $\str{M} =\<W,S,E,\in,(P_i)\>$ where 
$S\subseteq\wp(W)$ and $\in$ is the actual membership relation.
In the following we shall therefore also specify relational
models by just $\str{M}=\< W, S, E,(P_i)\>$, when the fact that $S \subset
\wp(W)$ and the natural interpretation of $\ee$ are understood.

Notice that a relational model $\str{M}$ induces a corresponding Kripke model $\str{K}(\str{M})$ on
$W$. We simply let $wRw'$ if for some $s\in S$ we have $wEs$ and
$w'\ee s$, and we let $R[w]:=\{w'\,|\,wRw'\}$. 

\subsection{Natural Classes of Relational Models}


\noindent In addition to extensionality and downward closure, we might
impose other constraints on a relational model $\str{M}$: in particular, we may require
$S$ to be the full powerset of $W$, or to resemble the powerset 
from the perspective of each world~$w$.

\smallskip
\begin{ddef}[classes of relational models]~\\
\label{relclassesdef}
A relational model $\str{M}=\<W,S,E,(P_i)\>$ is called:
\begin{itemize}
\item \emph{full} if $S=\wp(W)$;
\item \emph{locally full} if $S\supseteq\wp(R[w])$ 
for all $w\in W$.
\end{itemize}
\end{ddef}

\noindent
These conditions suggest different ways of encoding a concrete
inquisitive modal model $M=\<W,\Sigma,V\>$ as a relational model. 

\smallskip
\begin{ddef}[relational encodings]
\label{relencdef}
Let $\M=\<W,\Sigma,V\>$ be an inquisitive modal model. We define three 
relational encodings $\str{M}^{\ssc[\cdots ]}(\M)$ of $\M$, each based on~$W$, and with 
$wEs\Leftrightarrow s\in\Sigma(w)$, $w\;\ee\; s\Leftrightarrow w\!\in\! s$ and $P_i\!  =\! V(p_i)$. 
The encodings differ in the second sort domain~$S$:
\begin{itemize}
\item for $\MM^\rel(\M)$: $S=\mathrm{image}(\Sigma)$;
\item for $\MM^\lf(\M)$: $\;\,S=\{s\subseteq\sigma(w)\colon
  w\in W\}$;
\item for $\MM^\full(\M)$: $S=\wp(W)$.
\end{itemize}
\end{ddef}

\noindent
Clearly, $\MM^{\rel}(\M)$ is the minimal relational counterpart of~$\M$, $\MM^{\lf}(\M)$ its minimal counterpart that is locally full,
and $\MM^{\full}(M)$ its unique counterpart that is full.

\subsection{Relational Models and First-Order Logic}

\noindent
A relational inquisitive model supports a two-sorted 
first-order language having two relation symbols $\E$ and $\ee$, and a number of predicate symbols $\PP_i$ for $i\in I$. It is easy to translate formulae $\phi \in \inqbm$ to $\FO$-formulae 
$\phi^\ast(x)$ in a single free variable $x$ of the second sort in such a way that, if $\M$ is an inquisitive modal model and $\str{M}(\M)$ is any of the above encodings, we have:
\[
\M,s\models\phi \;\iff\; \MM(\M) \models \phi^\ast(s)
\]
This translation can be seen as an analogue of the standard
translation of modal logic to first-order logic. The framework of
relational inquisitive models thus allows us to view $\inqbm$ as a
syntactic fragment of $\FO$, $\inqbm \subset \FO$, just as standard
modal logic $\ML$ over Kripke structures 
may be regarded as a fragment $\ML\subset \FO$. 

Importantly, however, the class of relational inquisitive modal models
is not first-order definable in this framework, since the downward 
closure condition involves a second-order quantification. In other
words, we are dealing with first-order logic 
over non-elementary classes 
of intended models.

\section{ The $\sim$-invariant fragment of FO}
%
%
\label{inqbmsec}
\label{bisimconstraintsec}

\noindent
Regarding $\inqbm$ as a fragment of first-order logic  
(over relational models, in any one of the above classes),
we may think of downward closure and $\sim$-invariance as
characteristic semantic features of this fragment. The core question 
for the rest of this paper is to which extent $\inqbm$
may express all properties of worlds that are $\FO$-expressible.%
\footnote{Our results can be extended easily to address the 
corresponding question for properties of information states,
but due to space limitations we will not make the corresponding
results explicit here.} 
In other words, over which classes $\CC$ of models, if any, can
$\inqbm$  be characterised
 as the bisimulation invariant
fragment of first-order logic? In short, for what classes  $\CC$ do we have
\[
\inqbm \equiv \FO/{\sim} \quad \mbox{ $(\dagger)$}
\] 
just as $\ML \equiv \FO/{\sim}$ by van
Benthem's theorem?

\subsection{Bisimulation Invariance and Compactness} 

The inquisitive Ehrenfeucht--Fra\"\i ss\'e theorem, Theorem~\ref{EFthm}, 
implies $\sim$-invariance for all of $\inqbm$. By
Corollary~\ref{EFcorrworldpointed} it further implies \emph{expressive completeness}
of $\inqbm_n$ for any $\simn$-invariant property of world-pointed
models. 
In order to prove $(\dagger)$ in restriction to some particular class $\CC$ of
relational inquisitive models, 
it is thus necessary and sufficient to show that, for any $\phi(x) \in
\FO$, $\sim$-invariance of $\phi(x)$ over $\CC$ implies 
 $\simn$-invariance of $\phi(x)$ over $\CC$ for some finite $n$. 
This may be viewed as a \emph{compactness principle} for 
$\sim$-invariance of first-order properties, which is non-trivial in
the non-elementary setting of relational
inquisitive models.

\bO
\label{compactobs}
For any class $\CC$ of 
relational inquisitive models, the following are equivalent:
\bre
\item
$\inqbm \equiv \FO/{\sim}$ for world properties over $\CC$;
\item
for $\FO$-properties of world-pointed models, 
$\sim$-invariance over $\CC$ implies $\simn$-invariance over \CC\ 
for some~$n$.
\ere
\eO


Interestingly, first-order logic does not satisfy compactness in
restriction to the (non-elementary) class of relational inquisitive
models (see Example~\ref{compfailrelex} in the appendix).
More importantly, over the class of \emph{full} relational
models, violations of compactness can even be exhibited for 
$\sim$-invariant formulae.


\bO
\label{failcompobs}
Over full relational inquisitive models, the absence of infinite $R$-paths from the 
designated world $w$
(i.e., well-foundedness of the converse of $R$ at $w$)
is a first-order definable  
and $\sim$-invariant property of worlds that is not
preserved under $\simn$ for any $n$, hence not expressible in $\inqbm$. In
particular, first-order logic violates compactness over full relational models.
\eO

\subsection{The Characterisation Theorem}

\noindent
In light of Observation~\ref{compactobs}, Observation~\ref{failcompobs}
means that  $(\dagger)$ fails over the class of full relational models.
This is not too surprising: on full relational models,
$\FO$ has access to full-fledged second-order quantification, while
$\inqbm$ can only quantify over subsets within the range of~$\Sigma$.
This is in sharp contrast with our main theorem:

\medskip
\textsc{Theorem 2}. 
\emph{Let $\CC$ be either of the following classes of relational models: 
the class of all models; of finite models; of locally full models; of 
 finite locally full models. Over each of these classes, $\inqbm
 \equiv \FO/{\sim}$, i.e., a property of world-pointed models is definable
in \inqbm\ over $\CC$ if and only if it is both $\FO$-definable over $\CC$ 
and $\sim$-invariant over  $\CC$.}

\medskip
Without recourse to compactness, 
the most useful tool from first-order model theory for our purposes 
is the \emph{local nature} of first-order logic over relational structures,
in terms of Gaifman distance. 
In the setting of a relational model, Gaifman distance is graph distance in the
undirected bi-partite graph on the sets $W$ of worlds and $S$ of
states with edges between any pair linked by $E$ or $\ee$; the
\emph{$\ell$-neighbourhood} $N^\ell(w)$ of a world $w$ consists 
of all worlds or states at distance up to $\ell$ from $w$ in this
sense. 
It is easy to see that if $\str{M},w$ is a world-pointed relational model 
and $\ell\neq 0$ is even, the restriction of this model to $N^\ell(w)$, denoted  
$\str{M}\!\restr\! N^\ell(w),w$, is also a world-pointed relational
model.

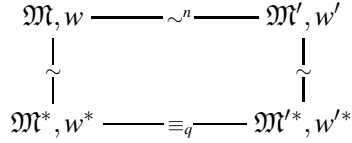
\begin{figure}
\[
\xymatrix{
\str{M},w \ar@{-}[d]|{\rule{0pt}{1ex}\sim} \ar @{-}[rr]|{\;\simn\,}
&& \str{M}',w' \ar @{-}[d]|{\rule{0pt}{1ex}\sim}
\\
\str{M}^\ast,w^\ast \ar@{-}[rr]|{\rule{0pt}{1.5ex}\;\equiv_q}
&& \str{M}'\nt^\ast,w'\nt^\ast 
}
\]
\caption{Generic upgrading pattern.}
\label{genupgradefigure}
\end{figure}

In light of Observation~\ref{compactobs}, to show that $(\dagger)$ holds over a class $\CC$ we need to 
show that a first-order formula $\phi(x)$ whose 
semantics is invariant under $\sim$ over the class $\CC$, 
is in fact invariant under 
one of the much coarser finite approximations $\simn$ 
over $\CC$, for some value $n$ depending on $\phi$.
For this there is a general approach that 
has been successful in a number of similar
investigations, starting from an elementary and constructive proof
in~\cite{OttoNote} of van~Benthem's
classical characterisation of basic modal logic~\cite{Benthem83} and
its finite model theory version due to Rosen~\cite{Rosen} (for ramifications of this method, 
see also~\cite{OttoAPAL04,DawarOttoAPAL09} and~\cite{Otto12JACM}).  
This approach involves an \emph{upgrading} of a sufficiently high 
finite level $\simn$ of bisimulation equivalence (or
$\equiv^n_\inqbm$)
to a finite target level $\equiv_q$ of  
elementary equivalence, where $q$ is the quantifier rank of $\phi$. 
Concretely, this  amounts to finding, for any world-pointed relational model 
$\str{M},w$, a fully bisimilar pointed model $\str{M}^\ast,w^\ast$ with the property 
that, if $\str{M},w\;\simn\;\str{M}',w'$, then  $\str{M}^\ast,w^\ast\equiv_q\str{M}'^\ast,w'^\ast$.
The diagram in Figure~\ref{genupgradefigure} shows how $\sim$-invariance of 
$\phi$, together with its nature as a first-order formula 
of quantifier rank $q$, entails its
$\simn$-invariance ---simply by taking the detour via the lower rung.
In the following section, we show how to achieve the required upgradings for various
classes $\CC$ of relational models; we use a variation on an upgrading technique 
from \cite{OttoNote}, based on an inquisitive analogue of partial tree unfoldings.

\subsection{Partial Unfolding and Stratification}
\label{stratsec}

\noindent
Theorem~\ref{main1} boils down to the compactness
property expressed in Observation~\ref{compactobs} for the relevant
classes of relational models. To show this property we make use of a 
process of \emph{stratification}, comparable to tree-like unfoldings in standard modal logic.

\begin{ddef}
\label{treelikestratdef}
We say that a relational inquisitive model $\str{M}$ 
is \emph{stratified} if its two domains $W$ and $S$ consist of essentially disjoint\footnotemark\ 
strata $(W_i)_{i\in\N}$ and $(S_i)_{i\in\N}$ s.t.
\bre
\item
$W = \dot{\bigcup}\, W_{i}$ and 
$S \setminus \{\emptyset \}  = \dot{\bigcup}\, (S_{i}\setminus\{\emptyset\})$;%
\addtocounter{footnote}{-1}%
\footnote{The $S_i$ will share the trivial information state
  $\emptyset$, by extensionality and the downward closure requirement. 
  This is unproblematic for our purposes. }
\item $S_i \subset \wp(W_{i+1})$, and  $E[w] \subset S_{i}$
for all $w\in W_{i}$.
\ere
For an even number $\ell\neq 0$ and a world $w$, we say that $\str{M}$ 
is \emph{stratified to depth $\ell$ from $w$} 
if $\str{M}\restr N^\ell(w)$ is stratified.
\end{ddef}

It is not hard to see that any world-pointed relational inquisitive
model is bisimilar to one that is stratified. Moreover, for any even number $\ell\neq 0$,
a finite world-pointed relational model is bisimilar to one that is finite 
and stratified to depth $\ell$ from its distinguished world. 
If the original model is locally full, the process of partial unfolding
leading to an ($\ell$-)stratified model preserves local fullness. 

\bO
\label{stratcutoffobs} 
For relational models $\str{M}$ and $\str{M}'$ that are stratified 
to depth $\ell$ for some even $\ell\neq 0$, and for
$n \geq \ell/2$:
\[
\barr{@{}r@{\;\;}l@{}}
&\str{M}\restr N^\ell(w),w 
\;\simn\; \str{M}'\restr N^\ell(w'),w'
\\
\Rightarrow & 
\str{M}\restr N^\ell(w),w\,\sim\, \str{M}'\restr N^\ell(w'),w'.
\earr
\]
\eO

\noindent
This is because, 
due to stratification and cut-off, the $n$-round game 
exhausts all possibilities in the unbounded game.





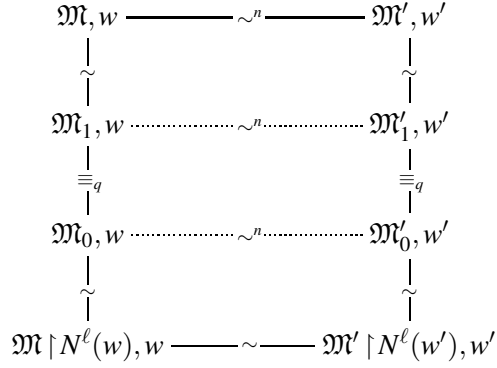
\begin{figure}
\[
\xymatrix{
\str{M},w \ar@{-}[d]|{\rule{0pt}{1ex}\sim} \ar @{-}[rr]|{\;\simn\,}
&& \str{M}',w' \ar @{-}[d]|{\rule{0pt}{1ex}\sim}
\\
\str{M}_1,w \ar@{-}[d]|{\rule{0pt}{1.2ex}\;\equiv_q}
\ar@{.}[rr]|{\;\simn\,}
&& \str{M}'_1,w'
\ar@{-}[d]|{\rule{0pt}{1.2ex}\;\equiv_q}
\\
\str{M}_0,w \ar@{-}[d]|{\rule{0pt}{1ex}\sim}
\ar@{.}[rr]|{\;\simn\,}
&& \str{M}'_0,w'
\ar@{-}[d]|{\rule{0pt}{1ex}\sim}
\\
\str{M}\restr N^\ell(w),w \ar@{-}[rr]|{\;\sim\,}
&& \str{M}'\restr N^\ell(w'),w' 
}
\]
\caption{Upgrading pattern for Theorem~\ref{main1}.}
\label{iqbmfigure}
\end{figure}

\medskip
\noindent
\textbf{Proof of Theorem~2.} 
Let $\CC$ be any one of the classes in the theorem 
and let $\phi(x) \in \FO_q$ be $\sim$-invariant over $\CC$. We want to
show that $\phi$ is $\simn$-invariant over $\CC$ for $n = 2^q$,
where $q$ is the quantifier rank of~$\phi$.
The upgrading argument is sketched in Figure~\ref{iqbmfigure}.
Towards its ingredients, consider 
a world-pointed relational model $\str{M},w$ 
in $\mathcal{C}$. Since $\phi$ is 
$\sim$-invariant, we can assume w.l.o.g.\ that $\str{M},w$ is
stratified to depth $\ell = n$. 
%
We define two world-pointed models $\str{M}_0,w$ and $\str{M}_1,w$ as follows. 
Both models contain $q$ distinct isomorphic copies
of $\str{M}$ as well as of $\str{M}\restr N^\ell(w)$. 
In addition, $\str{M}_0$ contains a copy 
of $\str{M}\restr N^\ell(w)$ with the distinguished world $w$,
 while $\str{M}_1$ contains a copy of $\str{M}$ with the 
 distinguished world~$w$:\footnote{To
   obey extensionality, when taking disjoint unions we identify the
   various copies of the empty information state coming from the
   different components of the model.} 
\[
\barr{@{}r@{\;:=\;\;}r@{\;\;\oplus\;\;}c@{\;\;\oplus\;\;}l@{}}
\str{M}_0,w &
q \otimes \str{M} &
\str{M}\restr N^\ell(w),w &
q \otimes \str{M}\restr N^\ell(w)
\\
\str{M}_1,w &
q \otimes \str{M} &
\str{M},w &
q \otimes \str{M}\restr N^\ell(w)
\earr
\]

\noindent
Using an Ehrenfeucht-Fra\"iss\'e game argument for $\FO$ 
it is possible to show 
(cf.\ Appendix~\ref{stratgameappendixsec} ) that
\[
\str{M}_0,w  \equiv_q \str{M}_1,w.
\]

Given any two pointed models $\str{M},w\,\simn\,\str{M}',w'$ in
$\CC$, we can see that $\phi$
is preserved between them by chasing the diagram in Figure~\ref{iqbmfigure}
along the path through the auxiliary models,  
which are all in $\mathcal{C}$.\hfill$\Box$

\section{Conclusion}


\noindent 
We have seen the foundations of a
model theory for inquisitive modal logic in two main aspects.
Firstly, the notion of \emph{inquisitive bisimulation equivalence} has been 
established as the appropriate notion of semantic invariance by an
Ehrenfeucht-Fra\"\i ss\'e correspondence, which provides a precious
tool 
for studying the expressive power of inquisitive modal logic.
Secondly, we have seen that $\inqbm$ admits
\emph{model-theoretic characterisations}
as the bisimulation-invariant fragment of classical first-order logic
over certain classes of relational structures with two sorts for
worlds and information states. 
Our result holds both in the general setting, and in restriction to finite models. 
The model-theoretic challenges arise in dealing with non-elementary 
classes of models, whose essentially two-sorted nature
extends first-order expressiveness in the direction of monadic 
second-order logic. Unpublished work~\cite{ICMOdraft17} 
indicates that this approach can be taken considerably further:
characterisations analogous to those presented here for basic  
$\inqbm$ can be obtained for \emph{inquisitive epistemic
  logic}---the multi-agent, S5-like variant of \inqml. 
In that setting, the model unfolding procedure that we used here to
establish our Theorem 2 can no longer be used, because the
resulting structures would no longer satisfy the inquisitive S5 constraints.
Instead, new and more complex techniques are needed.

%



\nocite{*}
\bibliographystyle{eptcs}
\bibliography{inqICMO}

\appendix

\section{Appendix}
\label{appendixsec} 

\subsection{Proof of Theorem~1}
\label{EFappendixsec}


\noindent
\textbf{Proof of Theorem~1/Proposition~\ref{prop:characteristic formulae}}.  
We fill in the missing part in the proof of Theorem~\ref{EFthm}, 
showing that the characteristic formulae, 
as defined in connection with Proposition~\ref{prop:characteristic formulae},
have the following properties:

\begin{enumerate}
\item $\M',w'\models\chi^n_{\M,w}\iff \M',w'\,\simn\, \M,w$
\item $\M',s'\models\chi^n_{\M,s} \iff \M',s'\,\simn\, \M,t$ for some  $t\subseteq s$ 
\item $\M',s'\models\chi^n_{\M,\Pi}\iff \M',s'\,\simn\, \M,s$ for some $s\in\Pi$
\end{enumerate}
First let us show that, if claim (1) holds for a certain $n\in\N$, then the claims (2) and (3) hold for $n$ as well. 

For claim (2), suppose $\M',s'\models\chi_{\M,s}^n$, that is, suppose $\M',s'\models\bigvee\{\chi_{\M,w}^n\,|\,w\in s\}$. This requires that for any $w'\in s'$ we have $\M',w'\models\chi_{\M,w}^n$ for some $w\in s$. By (1), this means that any world in $s'$ is $n$-bisimilar to some world in~$s$. Letting $t$ be the set of worlds in $s$ that are $n$-bisimilar to some world in $s'$, we have $t\subseteq s$ and  $\M',s'\sim_n \M,t$. 
Conversely, suppose $\M',s'\sim_n \M,t$ for some $t\subseteq s$. Then every $w'\in s'$ is $n$-bisimilar to some $w\in s$. By (1), this means that $\M',w'\models\chi_{\M,w}^n$, which implies $\M',w'\models\chi_{\M,s}^n$. Since this holds for any $w'\in s'$, and since $\chi_{\M,s}^n$ is a truth-conditional formula (by Proposition \ref{prop:truth-conditionality}), it follows that $\M',s'\models\chi_{\M,s}^n$. 

For claim (3), suppose $\M',s'\models\chi_{\M,\Pi}^n$. This implies 
$\M',s'\models\chi_{\M,s}^n$ for some $s\in\Pi$. By claim (2) we have $\M',s'\sim_n \M,t$ for some $t\subseteq s$. Since $\Pi$ is downward closed, $t\in\Pi$. Conversely, suppose $\M',s'\sim_n \M,t$ for some $t\in\Pi$. By (2), $\M',s'\models\chi_{\M,t}^n$, and since $t\in\Pi$, also $\M',s'\models\chi_{\M,\Pi}^n$.

Next, we use these facts to show that claim (1) holds for all
$n\in\mathbb{N}$,
by induction on $n$. 
The claim $\M',w'\models\chi_{\M,w}^0\Leftrightarrow\M',w'\!\sim_0\! \M,w $ follows immediately from the definition of $\chi_{\M,w}^0$. Now assume that claim (1), and thus also claims (2) and (3), hold for $n$, and let us consider the claim for $n+1$.

For the right-to-left direction,  suppose $\M',w'\sim_{n+1} \M,w$. We
want to show that $\M',w'\models\chi_{\M,w}^{n+1}$. This amounts to
showing that: (i) $\M',w'\models\chi_{\M,w}^{n}$; (ii)
$\M',w'\models\ibox\chi_{\M,\Sigma(w)}^{n}$; (iii)
$\M',w'\!\models\!\neg\ibox\chi^{n}_{\M,\Pi}$ when
$\Pi\subseteq\Sigma(w)$ and $\Pi\not\sim_n\Sigma(w)$. Let us show each
in turn.
\bre
\item $\M',w'\sim_{n+1} \M,w$ implies $\M',w'\sim_n \M,w$, so by the induction hypothesis $\M',w'\models\chi^n_{\M,w}$.
\item Take $s'\in\Sigma'(w')$. Since $\M',w'\sim_{n+1} \M,w$ we must have $\M',s'\sim_n \M,s$ for some  $s\in\Sigma(w)$. By the induction hypothesis, $\M',s'\models\chi_{\M,\Sigma(w)}^n$. This holds for all $s'\in\Sigma'(w')$, and so $\M',w'\models\ibox\chi^n_{\M,\Sigma(w)}$.
\item Suppose for a contradiction that for some
  $\Pi\subseteq\Sigma(w)$,  $\Pi\not\sim_n\Sigma(w)$  and
  $\M',w'\models\ibox\chi^n_{\M,\Pi}$. This means that every
  $s'\in\Sigma'(w')$ supports $\chi^n_{\M,\Pi}$ and thus, by our
  induction hypothesis, is $n$-bisimilar to some $s\in\Pi$. Since
  $\Pi\subseteq\Sigma(w)$ and $\Pi\not\sim_n\Sigma(w)$, there must be
  a state $t\in\Sigma(w)$ which is not $n$-bisimilar to any
  $s\in\Pi$. But since any state $s'\in\Sigma'(w')$ is $n$-bisimilar
  to some $s\in\Pi$, this means that $t$ is not $n$-bisimilar to any
  $s'\in \Sigma'(w')$. Since $t\in\Sigma(w)$, this contradicts the
  assumption that $\M',w'\sim_{n+1}\M,w$.
\ere

\noindent
This establishes the right-to-left direction of the claim. For the converse, suppose $\M',w'\models\chi^{n+1}_{\M,w}$. To prove $\M',w'\sim_{n+1}\M,w$, we must show that: (i) $w'$ and $w$ coincide on atomic formulae; (ii) any $s'\in\Sigma'(w')$ is $n$-bisimilar to some $s\in\Sigma(w)$; and (iii) any $s\in\Sigma(w)$ is $n$-bisimilar to some $s'\in\Sigma'(w')$. 
\bre
\item Since $\chi_{\M,w}^n$ is a conjunct of $\chi^{n+1}_{\M,w}$, by the induction hypothesis we have $\M',w'\sim_n \M,w$, which implies that $w$ and $w'$ make true the same atomic formulae.
\item Since $\ibox\chi^n_{\M,\Sigma(w)}$ is a conjunct of $\chi^{n+1}_{\M,w}$, $\M',w'\models\ibox\chi^n_{\M,\Sigma(w)}$. This implies that any $s'\in\Sigma'(w')$ supports $\chi^n_{\M,\Sigma(w)}$. By ind.\ hypothesis, this means that any $s'\in\Sigma'(w')$ is $n$-bisimilar to some $s\in\Sigma(w)$.
\item Let $\Pi$ be the set of states in $\Sigma(w)$ which are $n$-bisimilar to some $s'\in\Sigma'(w')$. Now, consider any $s'\in\Sigma'(w')$. We have already seen that $s'$ is $n$-bisimilar to some state $s\in\Sigma(w)$, which must then be in  $\Pi$ by definition. By  induction hypothesis, the fact that $s'$ is $n$-bisimilar to some state in $\Pi$ implies $\M',s'\models\chi^n_{\M,\Pi}$. And since this is true for each $s'\in\Sigma'(w')$, we have $\M',w'\models\ibox\chi^n_{\M,\Pi}$. 
Now suppose towards a contradiction that some $s\in\Sigma(w)$
were not $n$-bisimilar to any state in $\Sigma'(w')$. Then, $s$ would
not be $n$-bisimilar to any state in $\Pi$ either. This would mean
that $\Pi\not\sim_n\Sigma(w)$, which means that
$\neg\ibox\chi^n_{\M,\Pi}$ is a conjunct of $\chi^{n+1}_{\M,w}$. But
then, since $\M',w'\models\chi^{n+1}_{\M,w}$, we should have
$\M',w'\models\neg\ibox\chi^n_{\M,\Pi}$ contrary to what we found above. 
\ere
This completes the proof of Proposition~\ref{prop:characteristic formulae}. 
We can then use the properties of our characteristic formulae to prove the 
non-trivial direction of Theorem \ref{EFthm}. For suppose  
$\M,s\not\sim_n \M',s'$: then either of the states $s$ and $s'$ is not $n$-bisimilar to any subset of the other. Without loss of generality, say it is~$s'$. By the property of the formula $\chi^n_{\M,s}$ we have $\M,s\models\chi^n_{M,s}$ but $\M',s'\not\models\chi^n_{\M,s}$. Since the  modal depth of $\chi^n_{\M,s}$ is $n$, this shows that $\M,s\not\equiv^n_\inqbm\M',s'$.\hfill$\Box$

\subsection{Failures of Compactness} 
\label{compfailappendixsec}

Observation~\ref{failcompobs} refers to the following example of a 
first-order property of worlds $w$ in \emph{full} relational
models that is $\sim$-invariant and (as a well-foundedness assertion) 
obviously incompatible with compactness.
In terms of the accessibility relation 
$R = \{ (u,v) \colon v \in s \text{ for some $s$ with } uEs \}$:
\[
\mathbb{P}(w) :=\text{ there is no infinite }R\text{-path from }w\]


\noindent
This property is not expressible in $\inqbm$:
although it is $\sim$-invariant, it is clearly not invariant under 
any finite level of bisimulation equivalence. It is first-order
definable over full relational models, because those afford the
full expressive power of monadic second-order quantification over the
first sort, $W$, via first-order quantification over the second sort
$S = \wp(W)$. The following $\MSO$-formula, which defines $\mathbb{P}$ 
over the underlying Kripke frame, can therefore be
expressed in two-sorted first-order logic over full relational models:
\[
\neg \exists X \bigl(
x \in X \wedge \forall y \bigl( y \in X \rightarrow 
\exists z  (z \in X \wedge Ryz)\bigr)
\bigr).
\]

This shows that the analogue of our Theorem~2 fails for the class
of full relational models: over this class, there are properties that are $\FO$--definable
and $\sim$-invariant, but not definable in \inqbm.

It is also possible to show that compactness fails over the (non-elementary) class of all
relational models. 
However, in this case, Theorem~2 together with the compactness
of \inqbm\ implies that this cannot happen for $\FO$-formulae
that are $\sim$-invariant. 


The counterexample to compactness is obtained by relativising the
property $\mathbb{P}$ given above to a non-empty information
state $s$ in the image of $\Sigma$: in restriction to $s$, 
any relational encoding of $\<W,\Sigma\>$ provides the full power set 
$\wp(s)$ in the second sort.

\bE
\label{compfailrelex}
There is a first-order formula in a single free variable  
the second sort (information state) which over any relational
inquisitive model says of an element $s$ that (i) $s\in E[W]$ and (ii) 
there are no infinite $R$-paths included in $s$.
Condition (i) is expressed simply by $\exists w.Ews$; for (ii),
we relativise to $s$ the above formula that 
defines property $\mathbb{P}$, and then universally quantify
over $w \in s$.\hfill$\Box$
\eE

%

\subsection{Locality in Stratified Models} 
\label{stratgameappendixsec}

We present the game argument at the heart of the proof 
of Theorem~2 in Section~\ref{inqbmsec}, which needs to establish 
\[
(\ast\ast) \quad \str{M}_0,w  \equiv_q \str{M}_1,w
\]
for structures 
\[
\barr{@{}r@{\;=\;\;}r@{\;\;\oplus\;\;}c@{\;\;\oplus\;\;}l@{}}
\str{M}_0,w &
q \otimes \str{M} &
\str{M}\restr N^\ell(w),w &
q \otimes \str{M}\restr N^\ell(w)
\\
\str{M}_1,w &
q \otimes \str{M} &
\str{M},w &
q \otimes \str{M}\restr N^\ell(w)
\earr
\]
obtained as essentially disjoint sums 
from a world-pointed relational inquisitive model $\str{M},w$ 
that is stratified to depth $\ell = 2^q$, so that its truncation 
$\str{M}\restr N^\ell(w)$ is stratified. `Essentially disjoint sums'
here refers to disjoint sums with identifications of the empty
information states $\emptyset$ across the disjoint parts. It is not
hard to see that the $q$-equivalence claim in $(\ast\ast)$, however,
is insensitive to whether the empty information states, which are
uniformly present in the second sort of each component, 
are identified or not. So we may as well 
work with proper disjoint unions in the following proof.

\medskip\noindent
\textbf{Proof of $(\ast\ast)$}
We argue that the second player has
a winning strategy in the classical $q$-round Ehrenfeucht--Fra\"\i
ss\'e game over these two structures starting in the position with a 
single pebble on the distinguished world $w$ on either side. 
Indeed, the second player can force a win by maintaining
the following invariant w.r.t.\ the game positions 
$(\ubar;\ubar')$ for 
$\ubar = (u_0,u_1,\ldots, u_m)$ with $u_0 = w$ in $\str{M}_0$ and 
$\ubar' = (u'_0,u_1',\ldots, u_m')$ with $u'_0 = w$ in $\str{M}_1$ 
after round $m$, for $m=0,\ldots, q$, for $\ell_m := 2^{q-m}$:

\begin{quotation}\noindent
$\ubar$ and $\ubar'$ are partitioned into clusters of matching
sub-tuples such that the distance between separate clusters is greater
than $\ell_m$ and corresponding clusters are in isomorphic
configurations of isomorphic component structures of $\str{M}_0$ and
$\str{M}_1$ or in isomorphic configurations in
$\str{M}_0\restr N^\ell(w)$ and $\str{M}_1\restr N^\ell(w)$. 
\end{quotation}

\noindent
This condition is satisfied at the start of the game, for $m=0$.
The second player can maintain this condition 
through a round, say in the step from $m$ to $m+1$, as follows.
Suppose the first player puts a pebble in position $u= u_{m+1}$ in
$\str{M}_0$ or $u' = u'_{m+1}$ in $\str{M}_1$ at distance up to $\ell_{m+1}$ 
of one of the level $m$ clusters (it cannot fall within  distance
$\ell_{m+1}$ of two distinct clusters, since the distance between two
distinct clusters from the previous level is greater than $\ell_m = 2
\ell_{m+1}$); then this new position joins a sub-cluster of that cluster and 
its match is found in an isomorphic position relative to the matching cluster.
If the first player puts the new pebble in a position $u= u_{m+1}$ in
$\str{M}_0$ or $u' = u'_{m+1}$ in $\str{M}_1$ at distance greater 
than $\ell_{m+1}$ of each one of the level $m$ clusters, 
this position will form a new cluster and 
can be matched with an isomorphic position in one of the
as yet unused component structures on the opposite~side.

All steps of this proof  
restrict naturally to the
scenarios of (finite or general) locally full relational
inquisitive structures.

\end{document}